\documentclass[prl,twocolumn,nofootinbib]{revtex4-2}

\usepackage{graphicx}
\usepackage{color}
\usepackage{amsmath,amssymb}
\usepackage{fdsymbol}
\usepackage{bbold}
\usepackage{t1enc}

\usepackage{color}
\usepackage[dvipsnames]{xcolor}

\DeclareMathSymbol{\shm}{\mathbin}{AMSa}{"39}

\begin{document}

\title{$H \to ZZ$ as a double-slit experiment}
\author{J. A. Aguilar-Saavedra}
\affiliation{Instituto de F\'\i sica Te\'orica, IFT-UAM/CSIC, c/ Nicol\'as Cabrera 13-15, 28049 Madrid}

\begin{abstract}
The decay $H \to ZZ \to 4\ell$, with $\ell = e,\mu$, can be used to test quantum interference in analogy to the famous double-slit experiment. The observations corresponding to `covering a slit' can be extracted from data in the $ee\mu\mu$ channel, while the observations when both `slits' are open, which include the interference, are measured in the $4e/4\mu$ channels. In this way, this process offers a unique opportunity to investigate identical-particle effects at high energies, and provide the first evidence of identical-particle behaviour for muons. Sensitivity at the $4\sigma$ level could be achieved at the high-luminosity upgrade of the Large Hadron Collider.
\end{abstract}

\maketitle

Quantum mechanics (QM) is essential for our understanding of Nature. When performing tests of QM at high energies we are probing its boundaries and thereby extending the range in which it is known to be a correct description of Nature. At present there is no compelling alternative to extend QM as formulated in the early 20$^\text{th}$ century; however, this fact does not diminish the importance of performing tests at high energies and under extreme conditions, where QM might start to break down. 
Quantum mechanics underlies quantum field theory, which successfully describes all particle interactions via the Standard Model. Therefore, all production and decay processes observed at the Large Hadron Collider (LHC) are quantum in nature. Nonetheless, inherently quantum tests, such as those of entanglement already performed~\cite{ATLAS:2023fsd,CMS:2024pts,CMS:2024zkc}, remain of significant interest to probe the foundations of quantum mechanics at high energy.

In this Letter we consider the Higgs boson decay $H \to ZZ$ in the four-lepton final state, highlighting its role as a novel test of identical-particle effects and quantum interference. Tests of identical-particle effects have already been performed with photons~\cite{Hong:1987zz} and electrons~\cite{PhysRev.94.357}. The $H \to ZZ \to 4 \ell$ process with $\ell = e,\mu$ not only provides a further test for electrons at higher energies, but, remarkably, the first test for muons, for which experimental evidence of their identical-particle behaviour is lacking. 
Moreover, $H \to ZZ \to 4 \ell$ demonstrates quantum interference, in a way analogous to the classic double-slit experiment. 

There are several magnitudes that differ between the decays into identical and distinguishable fermions, for example, the total decay width and the invariant mass distributions. In addition to those, some angular correlations exhibit striking differences. We will  first set up our framework to study the angular correlations in the decay, to later discuss how to test interference and identical-particle effects.

{\em Framework.}
For the sake of clarity we first discuss the $ee\mu\mu$ final state, where the particles are distinguishable and the two intermediate bosons can be uniquely identified by the flavour of their decay products.
We will work in the helicity basis, a moving reference system with vectors $(\hat r, \hat n, \hat k)$ defined as follows~\cite{Bernreuther:2015yna}:
\begin{itemize}
\item $\hat k$ is taken in the direction of one of the $Z$ bosons in the Higgs rest frame; we select the one with largest invariant mass and label it as $Z_1$, with decay products $\ell_1^+$, $\ell_1^-$. The other boson is labelled as $Z_2$, with decay products $\ell_2^+$, $\ell_2^-$.
\item $\hat r$ is defined as $\hat r = \mathrm{sign}(\cos \theta) (\hat p_p - \cos \theta \hat k)/\sin \theta$, with $\hat p_p = (0,0,1)$ the direction of one proton in the laboratory frame, and $\cos \theta = \hat k \cdot \hat p_p$.% The definition for $\hat r$  is the same if we use the direction of the other proton $- \hat p_p$. 
\item $\hat n$ is taken orthogonal, $\hat n = \hat k \times \hat r$.
\end{itemize}
The angular orientation of the decay products can be specified by the polar and azimuthal angles $\Omega_1 = (\theta_1,\phi_1)$ and $\Omega_2 = (\theta_2,\phi_2)$ of the negative lepton momenta in the rest frame of the parent $Z$ boson, measured in the $(\hat r, \hat n, \hat k)$ reference system. The four-dimensional decay angular distribution reads~\cite{Aguilar-Saavedra:2022wam}
\begin{eqnarray}
\frac{1}{\sigma}\frac{d\sigma}{d\Omega_1d\Omega_2} & = & \frac{1}{(4\pi)^2}\left[ 1 +a_{LM}^1 Y_L^M(\Omega_1) + a_{LM}^2  Y_L^M(\Omega_2)   \right. \notag \\
& & \left. + c_{L_1 M_1 L_2 M_2} Y_{L_1}^{M_1}(\Omega_1)Y_{L_2}^{M_2}(\Omega_2)  \right] \,,
\label{ec:dist4D}
\end{eqnarray}
with $Y_L^M$ the usual spherical harmonics, and implicit sum over repeated indices. In the $ee\mu\mu$ final state with distinguishable particles the constants $a_{LM}^{1,2}$, $c_{L_1 M_1 L_2 M_2}$ are related to the coefficients in the spin density operator $\rho_{S_1 S_2}$ for the two $Z$ bosons~\cite{Aguilar-Saavedra:2022wam}. Specifically, this operator can be written as an expansion in irreducible tensors $T^L_M$,
\begin{eqnarray}
\rho_{S_1 S_2} & = & \frac{1}{9}\left[ \mathbb{1}_3 \otimes \mathbb{1}_3 + A^1_{LM} T^L_M \otimes \mathbb{1}_3 + A^2_{LM} \mathbb{1}_3\otimes T^L_M \right. \notag \\
& & \left. + C_{L_1 M_1 L_2 M_2}\ T^{L_1}_{M_1} \otimes T^{L_2}_{M_2} \right] \,,
\label{rho}
\end{eqnarray}
with constants $A_{LM}^{1,2}$, $C_{L_1 M_1 L_2 M_2}$. The precise definition of $T^L_M$ is not relevant here, and can be found e.g. in Ref.~\cite{Aguilar-Saavedra:2022wam}. For leptonic $Z$ decays one has $a_{LM}^{1,2} = B_L A_{LM}^{1,2} $, $c_{L_1 M_1 L_2 M_2} = B_{L_1} B_{L_2} C_{L_1 M_1 L_2 M_2}$, with $B_1 = - \sqrt{2\pi} \eta_\ell$, $B_2 = \sqrt{2\pi/5}$, and
\begin{equation}
\eta_\ell = \frac{1-4 s_W^2}{1-4 s_W^2 + 8 s_W^4} \simeq 0.13 \,,
\end{equation}
$s_W$ being the sine of the weak angle~\cite{Aguilar-Saavedra:2017zkn}. The $\eta_\ell$ factor suppresses the coefficients of spherical harmonics with $L=1$ in the distribution (\ref{ec:dist4D}), especially $c_{1010}$ and $c_{111-1}$.

In $4e / 4\mu$ final states with identical particles there are two possible pairings between opposite-sign leptons. (Each pairing corresponds to one of the two tree-level $H \to ZZ \to 4 \ell$ Feynman diagrams, related by exchange of identical particles.) Let us consider one such pairing and label with a common subindex the leptons in the same pair, $(\ell_1^+, \ell_1^-)$, $(\ell_2^+,\ell_2^-)$. One can then {\em define} the two $Z$ bosons by their momenta, $p_{Z_1} = p_{\ell_1^+} + p_{\ell_1^-}$, $p_{Z_2} = p_{\ell_2^+} + p_{\ell_2^-}$. And, as before, label as $Z_1$ the one with largest invariant mass. The reference system is then constructed as described for  distinguishable particles, and likewise the four-dimensional decay distribution has the form (\ref{ec:dist4D}).\footnote{In principle, higher-rank spherical harmonics may enter the angular distribution for $4e/4\mu$ final states. We have checked that the coefficients for $L=3$ are compatible with zero within Monte Carlo uncertainty (for several million events). This can be understood since a fermion pair in s-wave has total angular momentum $J=0,1$, and behaves like a scalar or spin-1 particle.}
However, the numerical values of the coefficients $a_{LM}^{1,2}$ and $c_{L_1 M_1 L_2 M_2}$ differ from the $ee\mu\mu$ final state. In particular, $c_{111-1}$ and $c_{1010}$ are up to 50 times larger, depending on the phase space region. We will precisely use the value of these coefficients to demonstrate the effects of interference and identical particles.

{\em Probing quantum interference.}
In the $ee\mu\mu$ final state there is a `correct' pairing of opposite-sign leptons (by flavour), let us label it as `A'. But one can also consider a `wrong' pairing, labelled as `B', in which we define $p_{X_1} =  p_{e^+} + p_{\mu^-}$, $p_{X_2} = p_{\mu^+} + p_{e^-}$. Neither $X_1$ nor $X_2$ correspond to any intermediate $Z$ boson in this final state, but $p_{X_i}$ and $p_{X_2}$ are time-like momenta, so we can boost the $\mu^-$ and $e^-$ momenta to the $X_1$ and $X_2$ rest frames, respectively, define the angles $\Omega_{1,2}$ and from the distribution (1) calculate coefficients $c_{111-1}$ and $c_{1010}$ --- which admittedly do not have any interpretation in terms of spin. (The motivation to consider this `wrong' pairing is the comparison with the $4e/4\mu$ final state, as we see below.)
Furthermore, we can perform the measurement of $c_{111-1}$ and $c_{1010}$ counting each event twice, one for each pairing. In this case, the result obtained is the average of the results obtained for each pairing, i.e. an incoherent sum.

In $4e/4\mu$ final states there is no `correct' pairing but we can still count each event twice, one for each pairing. And the results obtained for $c_{111-1}$ and $c_{1010}$ are {\em not} the average of pairings A and B as obtained for the $ee\mu\mu$ channel. 
In other words, the observation for $4e/4\mu$ is not the incoherent sum of the observations for the two choices in $ee\mu\mu$. (That would be the case in the absence of interference, as it can be seen by symmetry arguments.)
In terms of Feynman diagrams, each pairing in this final state corresponds to a diagram; however, our definitions and method to verify the interference by comparison with the $ee\mu\mu$ final state do not depend on expanding the amplitude in terms of diagrams.

This discussion can be illustrated with a Monte Carlo calculation at the parton level in full phase space. We generate $H \to ZZ \to 4\ell$ at the leading order with {\scshape MadGraph}~\cite{Alwall:2014hca}. For $ee\mu\mu$ we obtain
\begin{align}
& \text{Pairing A} && c_{111-1} = 0.098 && c_{1010} = -0.059 \,, \notag \\
& \text{Pairing B} && c_{111-1} = -1.210 && c_{1010} = 14.244 \,, \notag \\
& \text{Both pairings} && c_{111-1} = -0.556 && c_{1010} = 7.093 \,.
\end{align}
As said, the result when counting each event twice is the average of pairings A and B. For  $4e/4\mu$ we have
\begin{align}
& \text{Both pairings} && c_{111-1} = -0.032 && c_{1010} = 6.232 \,,
\end{align}
which is not the average of pairings A and B, and exhibits the interference effect.

The analogy with the double-slit experiment follows since in $4e/4\mu$ final states each lepton pairing is a possibility (e.g. to reconstruct the virtual $Z$ bosons), which can be regarded as a `slit'. Physical results are the coherent sum of both, as it is evident from the amplitude expansion in Feynman diagrams. In order to make the interference effect manifest we have considered the coefficients $c_{111-1}$, $c_{1010}$ obtained when summing over both pairings, and compared with the same quantity for the $ee\mu\mu$ final state. Following the analogy, the result of `covering a slit', that is, the result obtained with only one pairing, can be experimentally measured in $ee\mu\mu$. And it is found that in $4e/4\mu$ the result is not the average of both, again in analogy with the double-slit experiment.

\begin{figure}[t]
\begin{center}
\begin{tabular}{c}
\includegraphics[height=5.5cm,clip=]{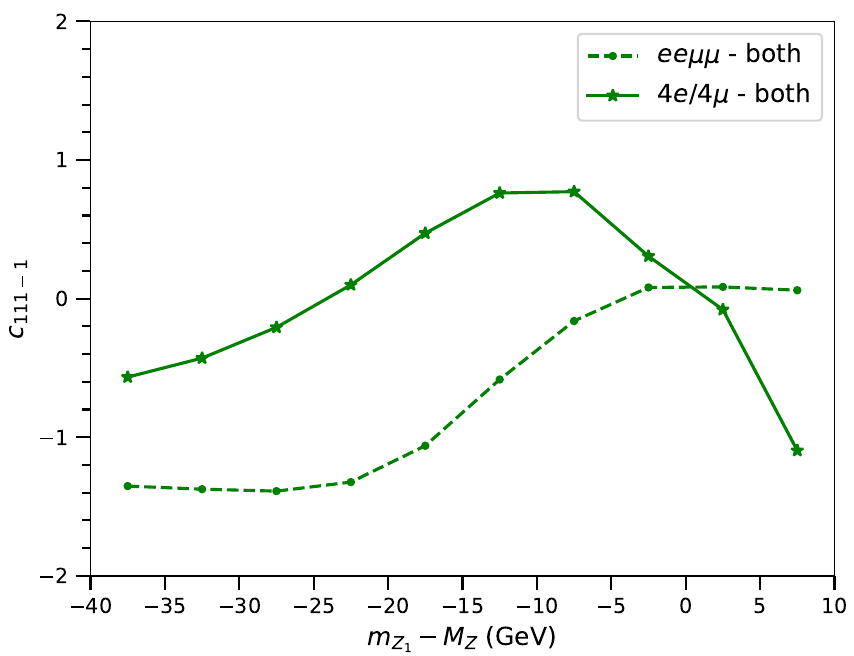} \\ 
\includegraphics[height=5.5cm,clip=]{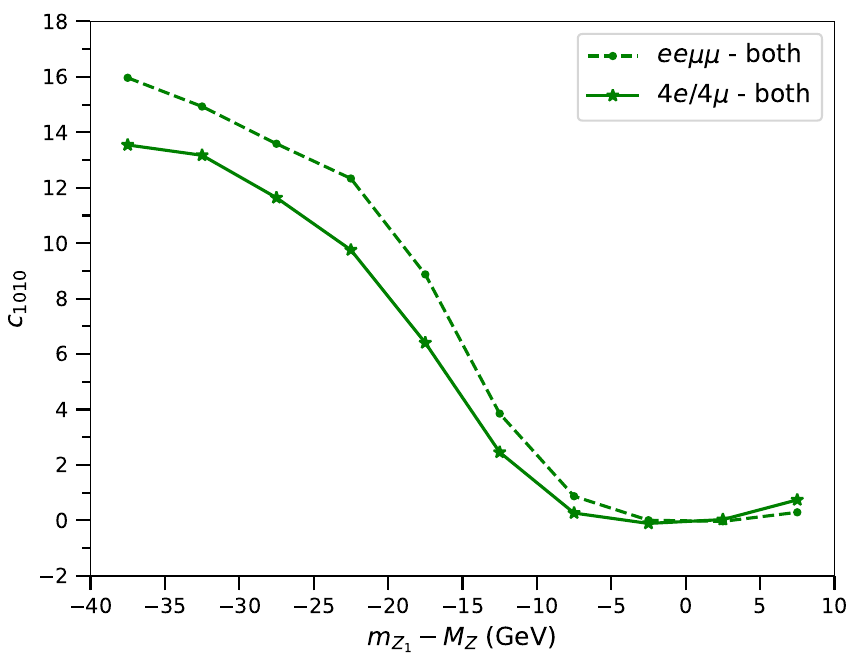} 
\end{tabular}
\caption{Dependence of the correlation coefficients $c_{111-1}$, $c_{1010}$ on $m_{Z_1}$, for $ee\mu\mu$ and $4e/4\mu$ final states, using both pairings of opposite-sign leptons. }
\label{fig:c2-M}
\end{center}
\end{figure}

The difference between the $4e/4\mu$ and $ee\mu\mu$ channels can also be studied as a function of $m_{Z_1}$, which is generally different for each pairing. The results are presented in Fig.~\ref{fig:c2-M}, in 5 GeV bins. Note that despite the fact that two pairings of a given event may in general lie in different bins, the differential distribution can be used as a probe of quantum interference.

{\em Probing identical-particle effects.}
In order to have a genuine test, one has to apply the same criterion to pair opposite-sign leptons in the $ee\mu\mu$ and $4e/4\mu$ final states and compare some observable, for example the values of $c_{111-1}$, $c_{1010}$ obtained using both pairings. We discuss here some alternatives.

\begin{figure}[t]
\begin{center}
\includegraphics[height=5.5cm,clip=]{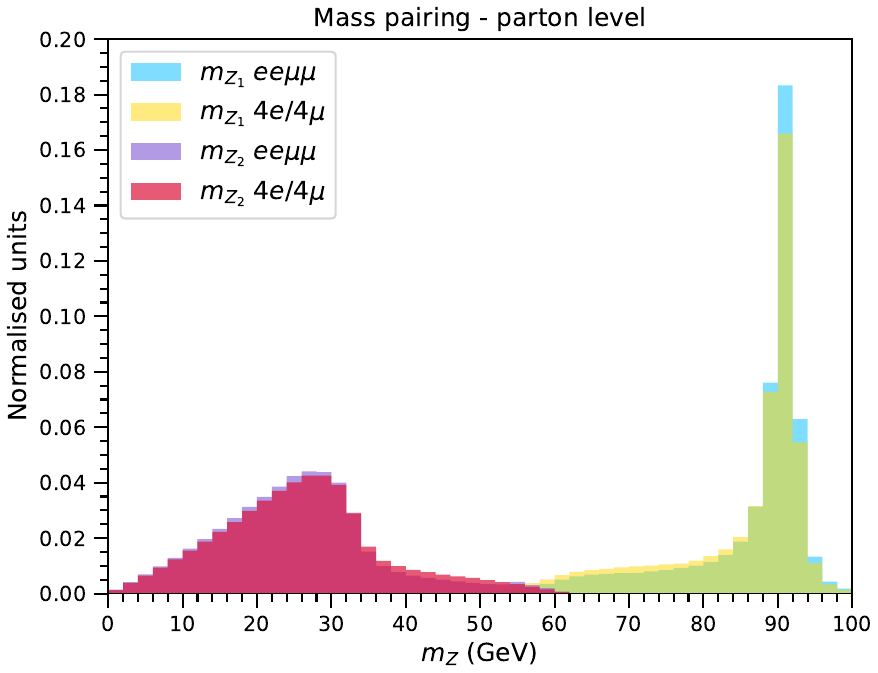} 
\caption{Kinematical distributions of $m_{Z_1}$ and $m_{Z_2}$, when applying mass pairing of opposite-sign leptons. }
\label{fig:M}
\end{center}
\end{figure}

\begin{figure}[t]
\begin{center}
\begin{tabular}{c}
\includegraphics[height=5.5cm,clip=]{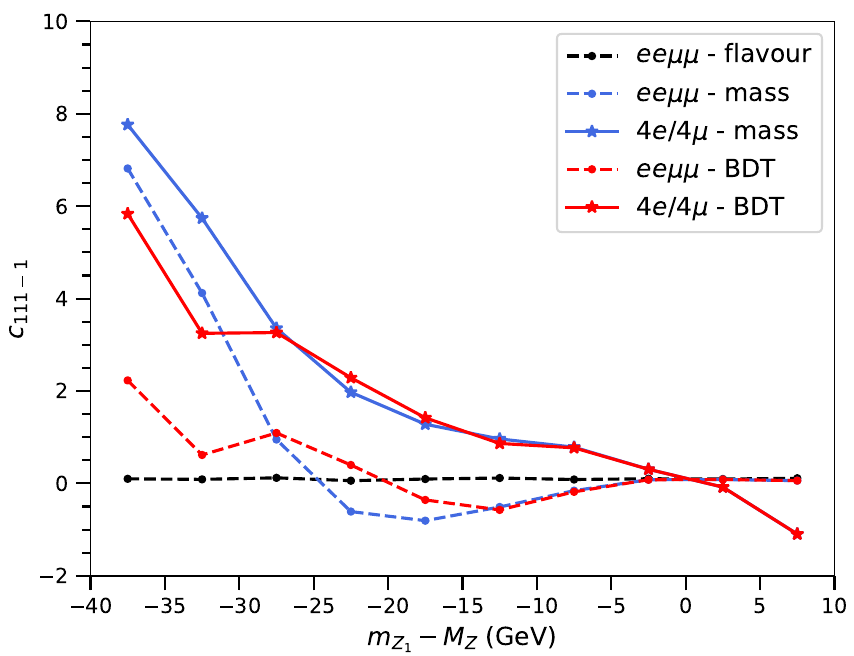} \\ 
\includegraphics[height=5.5cm,clip=]{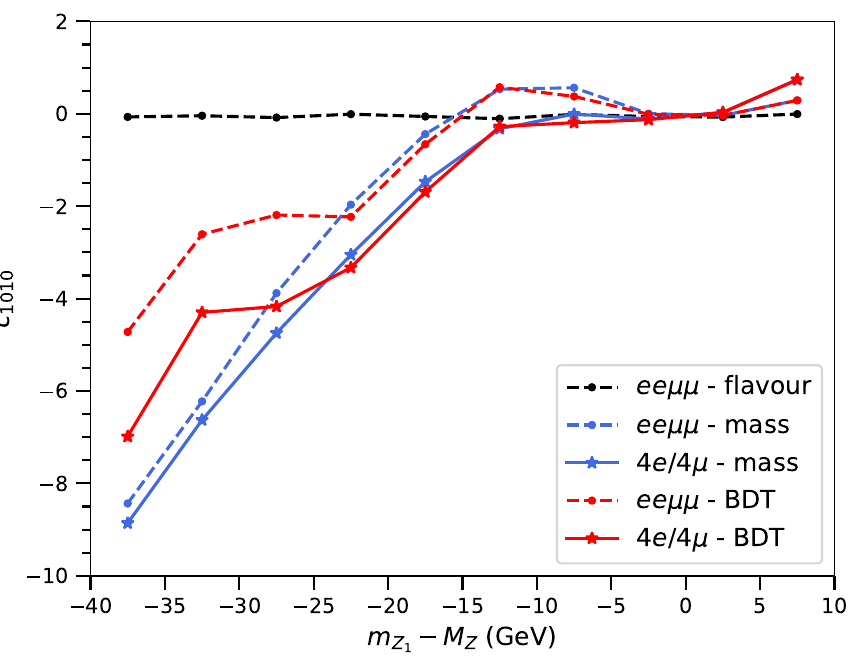} 
\end{tabular}
\caption{Dependence of the correlation coefficients $c_{111-1}$ and $c_{1010}$ on $m_{Z_1}$, using different methods to pair opposite-sign leptons. }
\label{fig:c-M}
\end{center}
\end{figure}

One simple but effective flavour-blind criterion that selects the correct pairing for $ee\mu\mu$ in the vast majority of the events is, among the two possibilities, to choose the one yielding the largest value of $m_{Z_1}$. We will denote this criterion as `mass pairing'. For low values of $m_{Z_1}$, i.e. when both bosons in $H \to ZZ \to ee\mu\mu$ are far from their mass shell, this criterion often fails to select the correct pairing; nevertheless, this phase space region amounts to a small fraction of the decay width.
The kinematical distributions of $m_{Z_1}$ and $m_{Z_2}$ are presented in Fig.~\ref{fig:M}. The former already provides a good discriminant to test identical-particle effects.
To improve the identification in the low-$m_{Z_1}$ region a boosted decision tree (BDT) can be trained using as input variables, for each of the two pairings (i) the two invariant masses of opposite-sign lepton pairs, $m_{Z_1}$, $m_{Z_2}$;
(ii) the cosines of the angles between leptons in the same pair, $z_1$, $z_2$, taking the momenta in the Higgs boson rest frame.
We implement this BDT using {\scshape XGboost}~\cite{Chen:2016:XST:2939672.2939785} with a maximum tree depth of 4, and a maximum of 100 trees, and otherwise default settings. The BDT is trained on the $ee\mu\mu$ final state. For $m_{Z_1} \geq 80$ GeV the BDT identifies the correct pairing in more than 99\% of the events. The percentage decreases to 85\% for $70 \leq m_{Z_1} \leq 75$ GeV, and drops to 58\% for $55 \leq m_{Z_1} \leq 60$ GeV. We find that a neural network obtains the same accuracy in the classification but with a much larger computing time. The distributions of $m_{Z_1}$ and $m_{Z_2}$ obtained with the BDT pairing are nearly identical to those in Fig.~\ref{fig:M}. The correlation coefficients for $ee\mu\mu$ are
\begin{align}
& \text{Mass pairing} && c_{111-1} = 0.087 && c_{1010} = -0.274 \,, \notag \\
& \text{BDT pairing} && c_{111-1} = 0.078 && c_{1010} = -0.200 \,,
\end{align}
and for $4e/4\mu$
\begin{align}
& \text{Mass pairing} && c_{111-1} = 0.758 && c_{1010} = -0.762 \,, \notag \\
& \text{BDT pairing} && c_{111-1} = 0.704 && c_{1010} = -0.715 \,.
\end{align}
Both coefficients are presented in Fig.~\ref{fig:c-M} as a function of $m_{Z_1}$ divided in 5 GeV bins.
For the $ee\mu\mu$ final state, the black lines correspond to the correct lepton pairing. Blue lines correspond to the mass pairing, and red lines are obtained with the BDT. In all cases, for $m_{Z_1} \geq M_Z - 20$ GeV, the mass and BDT pairing practically give the same results. For the $ee\mu\mu$ final state (dashed lines) we observe that both mass and BDT pairing give results for $c_{111-1}$ and $c_{1010}$ that are quite close to the small values obtained with the true pairing. Because the BDT and simple mass pairing yield quite similar discrimination, for simplicity we use the latter in the following.

{\em Future prospects.}
We estimate the statistical sensitivity of these measurements at the high-luminosity upgrade of the LHC (HL-LHC), with a luminosity of 3 ab$^{-1}$. We consider $gg \to H \to ZZ \to 4 \ell$ at 14 TeV, normalising the Higgs cross section to 54.67 pb~\cite{Cepeda:2019klc} and using branching ratios to $ee\mu\mu$ and $4e/4\mu$ of $5.9 \times 10^{-5}$ and $6.5 \times 10^{-5}$, respectively~\cite{LHCHiggsCrossSectionWorkingGroup:2016ypw}. Hadronisation and parton showering are performed with {\scshape Pythia}~\cite{Sjostrand:2007gs} and detector simulation with {\scshape Delphes}~\cite{deFavereau:2013fsa} using the default card for the HL-LHC. Electrons are selected in the pseudorapidity region $|\eta| \leq 2.5$, and with transverse momentum $p_T \geq 7$ GeV. Muons are selected with $|\eta| \leq 2.7$ and $p_T \geq 5$ GeV. We require two opposite-sign same-flavour leptons. The three higher-$p_T$ leptons must satisfy minimum thresholds of 20, 15 and 10 GeV, respectively. These event selection requirements are quite close to those in Ref.~\cite{ATLAS:2023mqy}. In addition, we select a window $120 \leq m_{4\ell} \leq 130$ GeV for the four-lepton invariant mass. In this region, the background has a size around 1/4 of the signal~\cite{ATLAS:2023mqy,CMS:2021ugl}. The expected number of events in the $ee\mu\mu$, $4e$ and $4\mu$ final states are 1144, 379 and 1106, respectively. At the reconstructed level, $c_{111-1}$ and $c_{1010}$ exhibit moderate deviations from the parton-level values. We collect them in Table~\ref{tab:creco}. For completeness we also give the values for the electroweak four-lepton background, calculated at the tree level, in the mass window $120 \leq m_{4\ell} \leq 130$ GeV.

\begin{table}[t]
\begin{center}
\begin{tabular}{cccccc}
&& \multicolumn{2}{c}{Both pairings} & \multicolumn{2}{c}{Mass pairing} \\
                                 & & $ee\mu\mu$ & $4e/4\mu$ & $ee\mu\mu$ & $4e/4\mu$  \\[1mm]
$c_{111-1}$ & ($H$)  & -0.76 & -0.03 & 0.05 & 0.92 \\
$c_{1010}$  & ($H$)  & 6.73  & 5.77  & -0.30 & -0.86 \\[2mm]
$c_{111-1}$ & (bkg)  & 1.45 & 1.42  & 0.42 & 0.42  \\
$c_{1010}$ & (bkg)  & 7.67 & 7.68  & -0.53 &  -0.69
\end{tabular}
\end{center}`
\caption{Correlation coefficients $c_{111-1}$, $c_{1010}$ at the reconstructed level, for the $H \to ZZ$ signal (`$H$') and the electroweak background (`bkg').}
\label{tab:creco}
\end{table}

The statistical uncertainty on $c_{111-1}$ and $c_{1010}$ is estimated by performing pseudo-experiments for each final state. In each pseudo-experiment, a subset of $N$ random events (the expected number of events for the final state considered) is drawn from the total sample, and for this subset the coefficients are determined from the distribution, without correction from detector-level to parton-level. A large number of $10^5$ pseudo-experiments is performed in order to obtain the probability density function (p.d.f.) of these quantities, and from it the statistical uncertainty. 
For the calculation of $c_{111-1}$, $c_{1010}$ we also apply an upper cut $m_{Z_1} \leq M_Z$. As it can be seen from Figs~\ref{fig:c2-M} and \ref{fig:c-M}, the region $m_{Z_1} \sim M_Z$ provides little discrimination, and in some cases there is a sign flip in the coefficients for $m_{Z_1} \geq M_Z$. 
No further sensitivity optimisation is attempted. Results are presented in Table~\ref{tab:c}. The statistical significance ($Z$-score) of the differences are given in the last column. The first value corresponds to the comparison of $ee\mu\mu$ and $4e/4\mu$ channels (summing both flavours) and the second one to the comparison between $ee\mu\mu$ and $4\mu$.

\begin{table}[h]
\begin{center}
\begin{tabular}{ccccc}
                    & $ee\mu\mu$        & $4e/4\mu$            & $4\mu$ & $Z$ \\[2mm]
$c_{111-1}$ & $-0.88 \pm 0.22$     & $-0.03 \pm 0.19$ & $-0.06 \pm 0.21$ & 2.9 | 2.7 \\
$c_{1010}$ & $7.68 \pm 0.27$       & $6.45 \pm 0.23$  & $6.47 \pm 0.27$  & 3.5 | 3.2 \\[2mm]

$c_{111-1}$ & $0.07 \pm 0.35$  & $1.20 \pm 0.30$   & $1.17 \pm 0.35$ &   2.4 |  2.2 \\
$c_{1010}$ & $-0.43 \pm 0.34$ & $-1.13 \pm 0.29$  & $-1.14 \pm 0.34$ &  1.6 |  1.5  \\
\end{tabular}
\end{center}
\caption{Central values and expected uncertainties for $c_{111-1}$ and $c_{1010}$ obtained from pseudo-experiments, in the different final states, and statistical significance of the differences (last column). The upper set of values is obtained with both pairings, and the lower one with mass pairing.}
\label{tab:c}
\end{table}

Quantum interference is probed by comparing the results obtained for the $ee\mu\mu$ and $4e/4\mu$ samples using both pairings (upper block of Table~\ref{tab:c}). The statistical differences for $c_{111-1}$ and $c_{1010}$ amount to $2.9\sigma$ and $3.5\sigma$, respectively. Their statistical correlation is -0.14, therefore both can easily be combined to obtain a significance of  $4.9\sigma$. (Without the upper cut on $m_{Z_1}$, the statistical differences are of $2.6\sigma$ and $2.9\sigma$, and the combination amounts to $4.2\sigma$.)
Identical-particle effects can be tested either using both pairings or mass pairing. The former gives more sensitive results. The comparison of $c_{111-1}$, $c_{1010}$ between $ee\mu\mu$ and $4\mu$ samples yields $2.7\sigma$ and $3.2\sigma$ which, when combined, provide an overall significance of $4.5\sigma$. 

Further tests of identical-particle effects can be performed by comparing the full $m_{Z_1}$ distributions for $ee\mu\mu$ and $4e/4\mu$ final states (or $4\mu$ alone) obtained with mass pairing. 
The $p$-value giving the probability that two distributions result from the same p.d.f. is obtained using a Kolmogorov-Smirnov test~\cite{K,S}. We perform $10^5$ pseudo-experiments, and for each one we calculate the $p$-value, which is then translated into a $Z$-score corresponding to the number of standard deviations. The mean value of the $Z$-score obtained over pseudo-experiments is the expected significance. For the difference between $ee\mu\mu$ and $4e/4\mu$ ($4\mu$) samples it amounts to $3.6\sigma$ ($3.2\sigma$). The distribution of $m_{Z_2}$ has a discriminating power below the $2\sigma$ level. 

A final comment on the electroweak four-lepton background is in order. In the region of interest it is quite small, as mentioned; the coefficients are also quite similar in $ee\mu\mu$ and $4e/4\mu$ final states, as seen in Table~\ref{tab:creco}. And they are of similar magnitude as the ones for $H \to ZZ$. Therefore, a background subtraction, which is out of the scope of the present work, seems feasible, and will only result in a mild increase of the statistical uncertainty in $c_{111-1}$ and $c_{1010}$, and a corresponding decrease of the significance with respect to the results presented here. Next-to-leading order corrections for the $ee\mu\mu$ final state are known~\cite{Grossi:2024jae}, and provide a small shift in $c_{111-1}$ and $c_{1010}$ which is below the statistical uncertainty.

{\em Discussion.}
In this Letter we have proposed a novel probe of quantum interference in Higgs decays $H \to ZZ \to 4\ell$, based on angular correlations of the decay products. This test only relies on a data comparison between $ee\mu\mu$ and $4e/4\mu$ final states. This is in contrast to other proposals; for example, the ALICE Collaboration has measured an angular oscillating pattern in Pb-Pb collisions~\cite{ALICE:2024ife} that can be interpreted as quantum interference but only after compared to Monte Carlo predictions. Whereas, in Higgs decays one can even measure the analogous to `covering a slit' to subsequently establish the presence of quantum interference by using data alone.

The measurements in Higgs decays can also be used to test identical-particle effects. To our knowledge, these would be the first tests to establish the identical-particle behaviour for muons. We have explored several observables that differ between $ee\mu\mu$ and $4e/4\mu$ final states, to conclude that the angular correlations used to probe quantum interference are also the most sensitive to establish the identical-particle behaviour. We have not considered the total decay width, whose difference between $ee\mu\mu$ and $4e/4\mu$ final states does not have a clear interpretation as due to identical-particle behaviour.

In our work we have focused, for simplicity, on the angular correlation coefficients $c_{111-1}$ and $c_{1010}$. Other terms in the distribution (\ref{ec:dist4D}) exhibit differences between $ee\mu\mu$ and $4e/4\mu$ final states, and they could also be incorporated in an experimental analysis to improve the statistical significance of the discrimination. For example, $a_{20}^{1,2}$ show differences at the $1\sigma-2\sigma$ level, while the differences in $c_{222-2}$, $c_{212-1}$ and $c_{2020}$ are below $1\sigma$.

Electroweak $ZZ$ production with both $Z$ bosons on their mass shell has a larger cross section than $H \to ZZ$, but is not useful to probe interference and identical-particle effects. The correlation coefficients are very similar in $ee\mu\mu$ and $4e/4\mu$ final states, with differences at the few percent level. This is due to the small contribution of the interference between diagrams related by identical-particle exchange in the latter. Below the $ZZ$ threshold, four-lepton production can be mediated by virtual $Z$ and $\gamma$ exchange; however, the differences between $ee\mu\mu$ and $4e/4\mu$ final states are small --- see for example Table~\ref{tab:creco} for the kinematical region near the Higgs peak. Then, $H \to ZZ \to 4\ell$, despite its small cross section, seems to be the most suited process to probe quantum interference and identical-particle effects, with an expected sensitivity at the $4\sigma$ level at the HL-LHC.

\section*{Acknowledgements}

I thank Y. Afik and A. Casas for useful comments.
This work has been supported by the Spanish Research Agency (Agencia Estatal de Investigaci\'on) through projects PID2022-142545NB-C21,  and CEX2020-001007-S funded by MCIN/AEI/10.13039/501100011033, and by Funda\c{c}{\~a}o para a Ci{\^e}ncia e a Tecnologia (FCT, Portugal) through the project CERN/FIS-PAR/0019/2021.

\end{document}